\documentclass[12pt]{article}

\usepackage{amsmath}
\usepackage{graphicx}
\usepackage{amsfonts}
\usepackage{amssymb}

\hbox{ \vrule height0pt width5in \vbox{\hbox{\rm }\break

\hbox{DF/UFJF-17/99 \hfill}
\hbox{UAB-FT-474 \hfill}
\hrule height.1cm width0pt}} \vspace{3mm}

\begin{document}

\begin{center}
\bigskip
\vskip1.3cm

{\Large On the scaling behavior of the cosmological 
constant and the possible
existence of new forces and new light degrees of freedom }

\vskip8mm
{\large Ilya L. Shapiro}$^{\,a,b}$ $\,,\,\,\,\,\,\,\,$
{\large  Joan Sol\`{a}}$^{\,c}$ 
\vskip5mm

\medskip

$^{a}$\textsl{Departamento de F\'{\i}sica, ICE, 
Universidade Federal de Juiz de Fora}

\textsl{Juiz de Fora, 36036-330, MG, Brazil}
\vskip 3mm

$^{b}$\textsl{Tomsk State Pedagogical University, Tomsk, Russia}
\vskip 3mm

$^{c}$\textsl{Grup de F\'{\i}sica Te\`{o}rica and Institut de F\'{\i}sica
d'Altes Energies, }

\textsl{Universitat Aut\`{o}noma de Barcelona, E-08193, Bellaterra, 
Barcelona, Catalonia, Spain}
\end{center}

\vspace{1.3cm}

\begin{center}
\textbf{ABSTRACT}
\end{center}

\begin{quotation}
\noindent{A large value of the cosmological constant (CC) is 
induced in the Standard Model (SM) of Elementary Particle Physics 
because of Spontaneous Symmetry Breaking. To provide a small value 
of the observable CC one has to introduce the vacuum term which 
cancels the induced one at some point in the very far infrared 
cosmic scale. Starting from this point we investigate
whether the cancellation is preserved at different energy scales. 
We find that the running of the Higgs mass, couplings and the vacuum 
term inevitably result in a scaling dependence of the observable 
value. As a consequence one meets a nonzero CC at an energy scale 
comparable to the typical electron neutrino mass suggested by 
some experiments, and the order of magnitude of this constant is
roughly the one derived from recent supernovae observations. However 
the sign of it is negative -- opposite to what is suggested by 
these observations. This discrepancy may be a hint of the existence 
of an extra very light scalar, perhaps a Cosmon-like dilaton, 
which should essentially decouple from the SM Lagrangian, but that 
it nevertheless could mediate new macroscopic forces in the 
submillimeter range. }

\end{quotation}

\newpage
\vskip 16mm

\section{Introduction}

According to the modern understanding, Particle Physics can be successfully
described by the Standard Model (SM) of the strong and electroweak
interactions, which could perhaps be approximately valid till some Grand
Unification scale $M_{X}\sim10^{16}GeV$ or even up to the Planck scale
$M_{P}\simeq10^{19}GeV$. The Spontaneous Symmetry Breaking (SSB) through the
Higgs mechanism is a crucial ingredient of the SM as it makes the weak gauge
vector fields massive while preserving the gauge invariance. 
As a by-product,
however, the introduction of a scalar Higgs potential and the SSB leads to a
non-vanishing value of the vacuum energy which can be interpreted as 
a direct
contribution to the cosmological constant (CC) in the Einstein-Hilbert 
action
for the gravitational field \cite{Wein1}:
\begin{equation}
S_{gr}
\,=\,-\,\int d^{4}x\sqrt{-g}\,\left(  \,\frac{1}{16\pi G}\;R\,+\,\Lambda
\,\right)  \,. 
\label{n6a}
\end{equation}
%%%%%%%%%%%%%%%%%%%%%%%%%%%%%%%%%%%%

Indeed, the SM contains a doublet of complex scalar fields $\Phi$. In the
ground (vacuum) state the expectation value (VEV) of $\,\Phi^{+}\Phi\,$ will
be denoted $\,<\Phi^{+}\Phi>\equiv\frac{1}{2}\phi^{2}$, where 
$\,\phi\,$ is a
classical scalar field. The corresponding classical potential reads
\begin{equation}
V_{cl}=-\frac{1}{2}m^{2}\phi^{2}+\frac{\lambda}{8}\phi^{4}. 
\label{2a}
\end{equation}
Shifting the original field $\phi\rightarrow H+v$ such that the physical
scalar field $H^{0}$ has zero VEV one obtains the physical mass of the Higgs
boson: $M_{H}=\sqrt{2}\;m$. Minimization of the potential (\ref{2a}) yields
the SSB relation:
\begin{equation}
\phi=\sqrt{\frac{2m^{2}}{\lambda}}\equiv v \,\,\,\,\,\,
\mathrm{and}
\,\,\,\,\,\,\, \lambda=\frac{\;M_{H}^{2}}{v^{2}}\,. 
\label{5N}
\end{equation}
The natural (experimental) electroweak mass scale in the SM is 
$\,M_{F}\equiv G_{F}^{-1/2}\simeq 293\,GeV$, 
where $G_{F\mathrm{{\ }}}=1.166\cdot10^{-5}
$\thinspace$GeV^{-2}$ is the physically measurable Fermi constant from muon
decay. The VEV of $\phi$ can be written entirely in terms of it:
$\,v=2^{-1/4}M_{F}\simeq 246\,GeV$. \ Similarly, from eq. (\ref{5N}) one
obtains the scalar coupling constant completely determined at that scale:
$\,\lambda=\sqrt{2}\;G_{F}\,M_{H}^{2}.$  All 
in all, even lacking at present
of direct experimental evidence of the Higgs boson, the indirect effects 
from
precision observables show that the SM is already a very much successful
Quantum Field Theory at the Fermi scale \cite{RADCOR98}. However, from
(\ref{2a}) 
and (\ref{5N}) one obtains the following induced value for the CC,
at the tree-level, that goes over to eq.(\ref{n6a}):
\begin{equation}
\Lambda_{ind}=<V_{cl}>=-\frac{m^{4}}{2\lambda}\,. \label{n6}
\end{equation}
If we apply the current numerical bound $\,M_{H}\gtrsim100\,GeV$ 
from LEP II,
then the value of $\left|  \Lambda_{ind}\right|  $ is $55$ orders 
of magnitude
larger than the observed upper bound for the CC -- typically this bound is
$\Lambda\lesssim10^{-47}GeV^{4}$ \cite{Wein1}. This is the Cosmological
Constant Problem in the SM of electroweak interactions.

To cure the CC problem one usually introduces a vacuum cosmological term
$\Lambda_{v}$ with opposite sign. The physical (observable) CC that finally
goes into eq.(\ref{n6a}) is then
\begin{equation}
\Lambda_{ph}=\Lambda_{ind}+\Lambda_{v}, \label{4}%
\end{equation}
and indeed the aforementioned astronomical upper bound applies only to this
quantity.

Some observations are in order. Since the consideration concerns gravity, it
would be natural to start from the renormalizable theory in curved 
space-time \cite{Book}. 
Such a theory contains the $\,\xi R\varphi^{2}\,$ term, and this
term seems 
to affect the SSB relations. This doesn't happen, however, unless
we imply a huge value of the parameter $\,\xi\,$. In order to see 
this, we
remind that the value of the curvature is defined (at least for small
energies, where higher order terms in the gravity action may be safely
omitted) from the Einstein equations, and therefore $\,R\sim8\pi
\,G\Lambda_{ph}\,$or $R\sim8\pi\,GT_{\mu}^{\mu}$. As $G=1/M_{P}^{2}%
\approx 10^{-38}GeV^{-2}$ 
and the observable density of the matter in the
Universe and current observational limits on the cosmological constant are
both of the same order, the actual value of the 
curvature scalar is very small
($R\sim10^{-84}GeV^{2})$ -- a 
reflex that a very small CC is equivalent to the
almost flatness of present day space-time--and  so 
the SSB relations do not
need to be altered. In fact, the relevant dimensionless parameter that
measures the deviation is $\,\xi R/m^{2}<\xi10^{-88}\ll1$ unless $\xi$ is
extremely large. Similar considerations apply at the Fermi epoch where the
picture remains unaltered unless $\xi\gg10^{33}$.

The introduction of the vacuum cosmological term is dictated also by the
requirement of renormalizability of the massive theory. Thus, the CC problem
is neither the existence of this term nor its renormalization, but the 
need of
the extremely precise choice of the corresponding normalization condition
$\Lambda_{ph}=0$ in the very far infrared (IR). 
At present, nobody knows the reason why the two 
terms on the right-hand side of (\ref{4}) -- plus perhaps some 
non-perturbative QCD contribution-- should
cancel each other with such an unnaturally big accuracy at that point.

%%%%%%%%%%%%%%%%%%%%%%%%%%%%%%%%%%%%%%%%%%%%%%%%%%%%%%%%%
%
\section{The running of the cosmological constant}
%%%%%%%%%%%%%%%%%%%%%%%%%%%%%%%%%%%%%%%%%%%%%%%%%%%%%%%%%

There were a number of attempts to solve the CC problem although 
no one of
them was successful enough \cite{Wein1}. In this letter we do not try to
understand the whys and wherefores of the original fine-tuning at very
low energies\footnote{Models have been suggested of finite theories 
in curved
space-time where $\Lambda$ naturally vanishes in the far IR. See e.g.
\cite{lam} and references therein. Their relation with the SM, though, 
is far
from clear.}, 
but rather the physical consequences of its quantum instability.
We address this problem by evaluating the quantum effects, namely the ones
that can be computed from the Renormalization Group (RG) method in the
framework of the effective field theory. It will be shown that the leading
contributions to the running of the physical CC (\ref{4}) cancel and that 
the
sub-leading contributions produce a negative CC at the energies 
above the mass
of the lightest fermion particle. This situation, however, can be very
different at higher energy domains near the Fermi scale $M_{F}$ or beyond
\cite{SS2}. In what follows, we are going to discuss the 
renormalization group
equations (RGE's) for quantities such as masses, couplings and CC. For the
investigation of the running of $\Lambda(\mu)$ -- the same concerns the
running of the gravitational constant $G,$ which will be studied elsewhere
\cite{SS2} -- one can always work with the beta-functions without taking 
into
account the classical dimensions for those quantities. The reason is that
these appear as powers of the dimensionless ratio 
$\,\,\left(  {\mu}/{\mu_{0}}\right)  .\,\,$ 
However the running constants acquire physical sense only
when they are substituted into Einstein equations. 
Since these equations are homogeneous in dimensions,
the previous ratio cancels automatically in them.

As we have already seen, the physical CC is nothing but the sum of 
two parts:
induced and 
vacuum, each of them satisfying its own RGE. Our initial framework
here will be just the SM, but later on we shall introduce some extra stuff.
The one-loop RGE for the vacuum part gains contributions from all massive
fields, and can be computed in a straightforward way by explicit 
evaluation of the vacuum bubble. 
In particular, the contribution from the
complex Higgs doublet $\Phi$ and the fermions is (for $\mu> M_{F}$)
\begin{equation}
(4\pi)^{2}\frac{d\Lambda_{v}}{dt}\,=\,2\,m^{4}\,-\,2\,\sum_{i}\,N_{i}
\,m_{i}^{4}
\,,\;\;\;\;\;\;\;\;\;\;\;\;\;\;\;\;\;\;\;\;\;\;\;\;\;\;\;\;\Lambda
_{v}(0)=\Lambda_{0}\,, 
\label{n11}
\end{equation}
where the sum is taken over all the fermions with masses $m_{i}$. Here
$\,t=\ln\mu$, and $N_{i}=1,3$ for leptons and quarks respectively.
$\Lambda_{v}$ has to be normalized at the very far IR cosmic scale, in order
to provide the precise cancellation of the induced contribution in 
(\ref{4}).

We shall study (\ref{n11}) and subsequent RGE's in some approximation. This
includes, in particular, omitting all higher loop contributions, and 
involves 
some application of the effective field theory approach. Indeed, this
approximation works better at lower energies. On the other hand, at 
these
energies the running of masses and couplings is weak, and therefore this
running can be disregarded together with the higher loops contributions. 
Thus,
to estimate the running of $\Lambda_{v}$, we assume that $m$ and 
$m_{i}$ take
fixed values characteristic of the Fermi scale.

On the other hand, the RGE for the induced CC follows, according to eq.
(\ref{n6}), from the general RGE's for the scalar mass $\,m\,$ and the
coupling $\lambda$ \cite{CEL}. In the SM, the latter read:
\begin{align}
&  (4\pi)^{2}\frac{dm^{2}}{dt}=m^{2}\; \left(  6\lambda-\frac{9}{2}
\;g^{2}-\frac{3}{2} \;g^{\prime}{}^{2}+2\,\sum_{i=q,l}\,N_{i}\,h_{i}
^{2}\right)  ,\;\;\;\;\;\;m^{2}(0)=m_{F}^{2}\,,\nonumber\\
&  (4\pi)^{2}\frac{d\lambda}{dt}= 12\lambda^{2}-9\lambda g^{2}-3\lambda
g^{\prime}{}^{2}+\frac{9}{4} \,g^{4}+\frac{3}{2}\,g^{2}g^{\prime}{}^{2}
+\frac{3}{4}\,g^{\prime}{}^{4}\,\nonumber\\
&  \hspace{2cm}+4\sum_{i=q,l}\,N_{i}\,h_{i}^{2}\left(  \lambda-h_{i}%
^{2}\right)  \hspace{2cm}\,\,\hspace{2.5cm}\,\lambda(0)=\lambda_{F}\,.
\label{n110}
\end{align}
Here $h_{l,q}$ are Yukawa coupling constants for the fermion 
fields. $\,q=(u,d,..,t)\,\,\,\,\mathrm{and}\,\,\,\,
l=(e,\mu,\tau,\nu_{e}
,\nu_{\tau},\nu_{\mu})\,$ label the type of spinor (quark and lepton) fields
of the SM. $\,\,\,g\,$ and $\,g^{\prime}\,$ 
are the $SU(2)_{L}$ and $U(1)_{Y}$ gauge couplings.
The boundary conditions for the renormalization group flow are imposed 
at the Fermi scale $M_{F}$ for all the parameters, with the important 
exception of $\Lambda_{v}$. Then, $\,\,g_{F}^{2}\approx0.4\,$ and 
$\,\;g^{\prime}{}_{F}^{2}\approx0.12\,$. 
In considering the equation for $\,\Lambda_{ind}\,$ we shall 
follow the same strategy as described above for eq. (\ref{n11}) -- that
is, we shall disregard the second order effects related to the running of
other parameters and attribute to all couplings and masses their constant
values at the Fermi scale.

A crucial point concerning the RGE's is the energy scale where they 
actually apply. 
This is especially important in dealing with the CC
Problem, since this problem is not seen at the Fermi epoch, but at 
the present
epoch, i.e. at energies far below the Standard Model scale $M_{F}$. The
corresponding beta-functions 
$\,\,\beta_{\Lambda_{v}},\beta_{m},\beta_{h_{i}},
\beta_{\lambda}\,...\,$ governing the evolution of the RGE's in a MS-like
scheme depend on the number of active degrees of freedom. These are the 
number
of fields 
whose associated particles have a mass below the energy scale $\mu$
that we are considering, because at sufficiently small energies one 
can invoke
the decoupling of the heavier degrees of freedom \cite{AC}. 
In this work we
are interested in the scaling behavior of the physical $\,\Lambda_{ph}$ at
very low energies. The importance of this energy domain is due to a recent
analysis of a set 
of high-redshift supernovae of Type Ia \cite{Supernovae}.
This analysis concludes that $\Lambda_{ph}>0$ at the $99\%$ 
$C.L.$ and it also
pinpoints a value 
for $\Lambda_{ph}$ of the order of a few times the matter
density $\rho_{m}$. Specifically, from the combined use of these data 
and CMBR
measurements \cite{CMBR} the following ranges are singled out: $ \Omega
_{m}=0.24\pm0.10$ and $\Omega_{\Lambda}=0.62\pm0.16$, where $\Omega_{m}
=\rho_{m}/\rho_{c}$ and $\Omega_{\Lambda}=\Lambda_{ph}/\rho_{c}$, with
$\rho_{c}=3H_{0}^{2}/8\pi G\simeq8.1\,h_{0}^{2}\times10^{-47}GeV^{4}$ the
critical density. Taking $h_{0}=0.65\pm0.10$ \cite{Silk} this gives, 
roughly,
$\Lambda_{ph}\simeq(2\pm1)\times10^{-47}GeV^{4}.$

%%%%%%%%%%%%%%%%%%%%%%%%%%%%%%%%%%%%%%%%%%%%%%%%%%%%%%%%%%%%%%%%
%%
\section{Cosmological constant and neutrino masses}
%%%%%%%%%%%%%%%%%%%%%%%%%%%%%%%%%%%%%%%%%%%%%%%%%%%%%%%%%%%%%%%%

In view of the previous consideration one can put forward the 
possibility that
this positive CC 
may be an effect of quantum scaling at an energy scale $\mu$
above the very far IR where $\,\Lambda_{ph}$ is zero. Following this 
line of thought, 
we may expect that the lightest degrees of freedom of the SM, namely
the neutrinos, are the only ones involved to determine $\Lambda_{ph}$ at
nearby IR points where we perform our measurements. In fact, since the
standard 
solution of the CC problem supposes an extremely exact fine tuning in
(\ref{4}), and since both vacuum and induced terms satisfy their own 
RGE's, it shouldn't be a great surprise if their exact cancellation 
breaks down by the
running. However, one may fear that this breaking can have a disastrous
effect. In the sense that, being both parts of (\ref{4}) many orders of
magnitude greater than their sum, any violation of the fine tuning might
produce a huge CC at any neighboring IR scale, a fact which would be 
blatantly inconsistent with cosmology. Indeed this could make all the 
approach based on a fine-tuning untenable.

Fortunately, nothing like that occurs. For the SM provides an automatic
cancellation mechanism between the running of $\,\Lambda_{ind}\,$ and
$\,\Lambda_{v}$, so that at very small energy scales only the second order
effects remain. 
These effects really drive $\,\Lambda_{ph}\,$ away from zero
(if we 
suppose that it started at zero value), but the order of magnitude of
this modification is small in the IR and can be consistent with 
the supernovae
observations. However, the sign of the resulting $\,\Lambda_{ph}$ 
is opposite
to the one derived from these observations, and one has to look 
for additional
physical input to cure this drawback.

To see all this in our RG approach, let us consider the energy region well
below the electron mass, where the only relevant particles are neutrinos
$\nu_{i}$. Then the effects from the other massive fields -- Higgs boson,
quarks, leptons and massive vectors -- in equations (\ref{n11}) and
(\ref{n110}) can be dropped. There are no contributions from photons and
gluons, so that we arrive at the following 
simplified equation for the vacuum term:
\begin{equation}
(4\pi)^{2}\frac{d\Lambda_{v}}{dt}\,=\,-\sum_{j}2\,m_{j}^{4}
\,,\;\;\;\;\;\;\;\;\;\;\;\; j=\nu_{e}\,,\nu_{\mu},\nu_{\tau}
\,,\;\;\;\;\;\;\;\;\;\;\;\; \Lambda_{v}(0)=\Lambda_{0}\,. \label{n19}%
\end{equation}
One can, indeed, 
rewrite the last equation in terms of the Yukawa couplings
$h_{j}$. We assume that neutrinos get masses through the 
usual mechanism in
the SM, and therefore $m_{j}=h_{j}v/\sqrt{2}$. Of course, the quantities
$h_{j}$ 
are very tiny, many orders smaller that any other coupling of the SM.

Next we explicitly check that the RGE for the induced counterpart
$\,\Lambda_{ind}\,$ exhibits (in the SM) an important cancellation, 
namely
that of the leading $\,m^{4}h_{j}^{2}/\lambda\,$ terms, 
and one is left with
$\, m^{4}h_{j}^{4}/\lambda^{2} $ which are much smaller for neutrinos.
 Indeed, from (\ref{n6}) and (\ref{n110})
\begin{equation}
\frac{d\Lambda_{ind}}{dt}\,= 
\,\frac{m^{4}}{2\lambda^{2}}\,\frac{d\lambda}
{dt}\,-\,\frac{m^{2}}{\lambda} \,\frac{dm^{2}}{dt}\,=
\,-\,\frac{1}{(4\pi)^{2}
}\cdot\frac{2m^{4}}{\lambda^{2}}\,\cdot\,\sum_{j}h_{j}^{4}\, =\,-\frac
{2}{(4\pi)^{2}} \sum_{j}m_{j}^{4}\;, \label{ccel1}
\end{equation}
where we have used the fact that in the SM the 
coefficient $\,\,m^{4}h_{j}
^{4}/\lambda^{2}\,\,$ is nothing but $m_{j}^{4}$. If the fermion masses
$m_{j}$ are very small, the quantum scaling evolution of the induced CC is
slow enough not to disturb 
the standard cosmological scenario. Moreover, at
very low energies, we see that the RGE's for the 
vacuum and induced CC are
identical and turn out to be very simple. Therefore the running of the
physical CC satisfies, in this energy domain, the equation
\begin{equation}
(4\pi)^{2}\frac{d\Lambda_{ph}}{dt}\,=
\,-4\sum_{j}m_{j}^{4},\;\;\;\;\;\;\;
\;\;\;\;\;\;\;\;\;\;\;\;\;\;\Lambda_{ph}(0)=
\Lambda_{ph}(IR). \label{newfor}
\end{equation}
Here $\Lambda_{ph}(IR)$ is the value of the physical CC in the very far
infrared, which we assumed $\Lambda_{ph}(IR)=0$. Therefore, 
according to our framework, a non-vanishing value of 
$\Lambda_{ph}$ is to be generated from the
lightest degrees of freedom available in the Universe.

As already mentioned, in the minimal SM only neutrinos could perhaps have
masses small enough such that their contribution to $\left|  \Lambda
_{ph}\right|  $ lies in the correct ballpark$.$ However, the spectrum of
masses and mixing angles predicted by different neutrino experiments
do entail some 
restrictions on the parameters \cite{DESY}. In particular, an
additional 
(sterile) neutrino $\nu_{s}$ is frequently invoked in many texture
models for the neutrino masses. 
(In the strict SM case, $\nu_{s}$ would simply
be a RH neutrino.) 
There are a variety of possibilities, but one can easily
check that in general 
not all neutrino species (if any) could be adequate to
generate 
a value for $\left|  \Lambda_{ph}\right|  $ in the desired range. In
fact, the pattern of 
neutrino masses must be such that it can resolve both the
Solar 
Neutrino Problem and the Atmospheric Neutrino Problem. One possible
mass 
hierarchy is that both the electron neutrino and the sterile neutrino are
much lighter than the rest: $m_{\nu_{e}},m_{\nu_{s}}\ll m_{\nu_{\mu}}
,m_{\nu_{\tau}}$.  The ANP solution would follow from 
$\nu_{\mu}-\nu_{\tau}$
oscillations near maximal 
mixing, and the SNP from taking the masses
$m_{\nu_{e}},m_{\nu_{s}}$ 
in the ballpark suggested by the  nonadiabatic MSW
mechanism (resonant oscillation inside the solar medium) -- 
leaving aside the
so-called vacuum oscillations solution, which is considered marginal
\cite{DESY}. The MSW effect allows the following two ranges: a 
region $\delta
m_{ex}^{2}=(0.3-1)\times10^{-5}GeV^{2}$ at small mixing angle, 
and a region of
$ \delta m_{ex}^{2}$ roughly between $(1.5\times10^{-5}-2\times
10^{-4})\,GeV^{2}$ at nearly maximum mixing angle. Specific models have 
been
suggested 
in the literature that satisfy these requirements \cite{MohapValle}.

In this framework, not only the SNP and ANP problems could be disposed 
of, but
it could also afford the necessary hot dark matter (HDM) component, provided
$m_{\nu_{\mu}},m_{\nu_{\tau}}$ are almost degenerate and in the $eV$ range.
This in turn could make the LSND result not that ``untenable'' with 
respect to
the remaining 
neutrino experiments \cite{DESY}. Therefore, we concentrate on
the ``double hierarchy Ansatz'' $ m_{\nu_{e}}\simeq m_{\nu_{s}}\ll
m_{\nu_{\mu}}\simeq m_{\nu_{\tau}}$, which could perhaps resolve 
the SNP, ANP,
HDM and LSND 
neutrino problems. We will see that it, too, might reveal itself
as the most suitable scenario to understand the present status of the
cosmological constant.

To start with, we point out that such an scenario is roughly compatible
with, say, $m_{\nu_{e}}$, $m_{\nu_{s}}$ =$(2-3)\times10^{-3}eV$ within the
MSW frame at 
small mixing angle; and so, by integrating (\ref{newfor}) in the
$\mu$ interval comprised between the masses $m_{\nu_{e}}$ and 
$m_{\nu_{s}}$
one easily finds $\left|  
\Lambda_{ph}\right|  \lesssim10^{-48}\,GeV^{4}$ i.e.
a number about one order of magnitude below the right range . On the other
hand, in the MSW region of maximal mixing between the electron and sterile
neutrinos one can amply achieve the desired 
$\left|  \Lambda_{ph}\right|  $.
 Notice that within the double hierarchy Ansatz the other two neutrino
species are $10^{2}-10^{3}$  times heavier, and hence do not enter the RG
analysis of $\Lambda_{ph}$ in the far IR. But of course they would enter at
earlier cosmological times -- and at some point also the remaining  SM
particles, which determine the evolution of the CC at the Fermi epoch
\cite{SS2}. Now, in spite of the fact that the last neutrino scenario is
compatible with the correct order of magnitude of $\Lambda_{ph}$, 
the sign of this 
parameter as predicted by eq. (\ref{newfor}) is negative, contrary to the
famous supernovae result \cite{Supernovae}.

%%%%%%%%%%%%%%%%%%%%%%%%%%%%%%%%%%%%%%%%%%%%%%%%%%%%%%%%%%%%%%%%%%
%%
\section{New light scalar fields and new forces}
%%%%%%%%%%%%%%%%%%%%%%%%%%%%%%%%%%%%%%%%%%%%%%%%%%%%%%%%%%%%%%%%%%

Intriguingly enough, the onerous sign from neutrinos could be remedied by
introducing an extra (real) light scalar field $S$ of non-vanishing mass
$m_{S}.$ We may also compute its contribution to the vacuum bubble. 
Then the
RGE for the physical cosmological constant becomes

\begin{equation}
(4\pi)^{2}\frac{d\Lambda_{ph}}{dt}\,=\beta_{\Lambda}\,\equiv\frac{1}{2}
\,m_{S}^{4}\,-\,4\,\sum_{i}\,m_{\nu_{i}}^{4};\;\;\; \;\Lambda_{ph}(0)=0\,,
\label{MasterRGE}
\end{equation}
where the sum extends over the electron neutrino and an sterile neutrino
according to the double hierarchy pattern mentioned above. For definiteness,
we may assume that  their average mass is $m_{\nu}\sim2\cdot10^{-3}eV.$

Let us briefly discuss possible ways to realize this program. Of course one
could invoke the existence of a ``just-so''\emph{\ } scalar field
\footnote{The existence of 
an \emph{ad hoc} scalar particle could be proposed only to
saturate the bound on $\Lambda$ \cite{Beane}, independently of neutrinos --
if
they are assumed massless. However, all hints seem to point towards light
massive neutrinos and $\Lambda>0$. Therefore, in a consistent picture 
these
must enter the game too.} with a mass of order $m_{S}\gtrsim2$ $m_{\nu}$ 
in
order to flip the sign of the overall $\beta_{\Lambda}-$function in
(\ref{MasterRGE}). In actual fact, $m_{S}$ must be a few times larger than
$m_{\nu}$ ($m_{S}\gtrsim4\,m_{\nu}$) 
in order to insure that after integrating
eq.(\ref{MasterRGE}) in the interval from $m_{\nu}$ to $m_{S}$ a value for
$\Lambda_{ph}$ obtains in the right ballpark. As for the interactions, 
this scalar should effectively be an ``sterile 
scalar'' -- very weekly interacting.
We recall that there are stringent phenomenological constraints on light
scalars that severely restrain their couplings and ranges \cite{gravimeter}.
In particular, a light Higgs boson is completely ruled out both by 
low-energy
and LEP experiments \cite{Hunter}.

On the other hand, an axion-like pseudoscalar is not precisely what we want
here. For, although axion mass windows presently constrain it to be in 
a mass
range whose upper bound is $ \sim10^{-3}eV$ \cite{gravimeter} -- 
and so still
of the order of the ultralight neutrinos mentioned above --, 
the problem with
axions in this context 
is that an essential relation like eq. (\ref{ccel1}) 
is
very peculiar to the Higgs structure of the SM. Therefore there is no
guarantee that it could be preserved -- without fine-tuning of 
the parameters
-- in the framework of general two-Higgs-doublet models. Similarly, we
understand that additional Higgs bosons are also used in some specific 
models
in the literature in order to implement the correct pattern of 
neutrino masses
mentioned above through radiative corrections\thinspace\cite{MohapValle}.
However, we do not commit ourselves to the underlying structure of these
models. In our framework the Higgs sector at the Fermi scale should
effectively behave as that of the SM.

Therefore, if we choose not to depart much from the RG structure of the SM,
we
have to resort to alternative scenarios. For example, we could entertain the
possibility that the necessary light scalar $S$ is a dilaton-like field. 
Such
a field can be identified to be the Goldstone boson that emerges from a
non-linearly realized  formulation of global dilatation-invariance
\cite{Coleman}. Namely, one assumes that the dilatation symmetry of the SM
Lagrangian is spontaneously broken at some high energy scale $M$. Then any
operator in the SM Lagrangian (including the vacuum term) is made 
invariant by
multiplying it by a suitable power of $e^{S/M}$. The scale variance of that
operator is compensated by a corresponding shift $S\rightarrow S+cM$
characteristic of a non-linearly realized Goldstone mode.

It is understood that a kinetic term for the non-linear field $S$ and the
Einstein piece (\ref{n6a}) also enter the total Lagrangian, and that the
gravitational 
part is made dilatation invariant following the same philosophy,
which in this case implies that the Ricci scalar $R$ is replaced by
$R\,e^{2S/M}.$ Notwithstanding, the scale symmetry of the SM plus Einstein
Lagrangians so constructed is explicitly broken by quantum effects. 
These give
rise to the trace anomaly, so that the Goldstone boson $S$ is not exactly
massless. The properties of this boson are very similar to the ones first
described 
in Ref. \cite{PSW} and for this reason we still dub it a ``Cosmon'',
because it may help us to understand the present status of the CC
problem
\footnote{For other developments and applications of the Cosmon model,
see \cite{CW,JSP}.}. It should, however, be clear that we adopt here a 
rather
different point of view than in the previous references. After all,
$\Lambda_{ph}$ was measured to be nonzero! Still, the present field $S$ has
similarly qualitative, though quantitatively different, phenomenological
consequences, as it will be shown below.  

The Cosmon mass, $m_{S},$ is determined by the interplay between the hard
scale defined by the trace anomaly and the scale $M$ of SSB of dilatation
symmetry. The anomalous trace of the energy-momentum tensor is given by 
the sum
\begin{equation}
\Theta_{\mu}^{\mu}=\frac{\beta(g_{s})}{2g_{s}}F_{a}^{\mu}F_{\mu}^{a}
+\Theta_{W}\,+\Theta_{G}, \label{Thetamumu}
\end{equation}
where the first term is the QCD part, $\Theta_{W}$ is the weak 
component
and $\Theta_{G}$ a possible gravitational contribution. The last 
two terms should be 
completely negligible at very low energies in our effective field
theory approach. Following \cite{PSW}, the Cosmon mass ensues from 
the equation of motion of $S,$ which is determined by the anomalous 
Ward identity generated by the coordinate covariant derivative of 
the dilatation current:
\begin{equation}
\nabla_{\mu}J^{\mu}=\Theta_{\mu}^{\mu}\,. \label{AWI}
\end{equation}
For instance, the scalar contribution to $J^{\mu}$ reads
\begin{equation}
J^{\mu}=\sqrt{-g}\left\{  (1+12h)Me^{2S/M}\partial^{\mu}S
+\Phi\partial^{\mu}\Phi\right\}  \,. 
\label{DilatJ}
\end{equation}
Notice that the 
particular piece in this dilatation current, namely the one
proportional to
\begin{equation}
h\equiv\frac{1}{16\pi}\,\frac{M_{P}^{2}}{M^{2}}\,, 
\label{vh}
\end{equation}
is a direct 
contribution from the dilatation-invariant scalar curvature term.
The linearized equation (\ref{AWI}) will become the field equation 
of motion
for $S$ if there is a stable value $S=S_{0}$ where 
$\,\left\langle \Theta_{\mu}^\mu\right\rangle\,$ vanishes. 
The $S$-dependence of the latter could 
appear e.g. through a change of gauge $\beta-$functions at a higher 
scale \cite{PSW}.
For $S_{0}/M\ll1,$ the structure of the SM will remain basically 
unaltered up
to terms of $ \mathcal{O}(1/M)$ which should be small if $M$ is 
sufficiently
large$.$ After expanding around $S=S_{0}$, and neglecting the
 electroweak and
gravitational contributions, simple dimensional analysis tells us that the
mass (squared) of $S$ is essentially given by
\begin{equation}
m_{S}^{2}=\frac{\Lambda_{QCD}^{4}}{M^{2}}\,, \label{MS}%
\end{equation}
where $\,\Lambda_{QCD} \sim 100\,MeV\,$ is the intrinsic QCD scale. 
Taking
$ M\gtrsim10^{10}GeV$  eq. (\ref{MS}) gives $m_{S}\lesssim\mathcal{O}
(10^{-3}-10^{-2})\,eV$  for the Cosmon mass\footnote{One could also obtain
this same value through e.g. $m_{S}^{2}=M_{F}^{4}\,/M^{2}$ by choosing $M$
near the more popular GUT scale $M_{X}=10^{16}GeV.$ However, it is 
difficult
to imagine how the electroweak dilatation anomaly could dominate in our
approach in the far IR.  So we do not expect the scale $M_{F}$ to be
involved. Moreover, at variance with the existence of an intrinsic,
RG-invariant, scale $\Lambda_{QCD}$ in QCD, there is no obvious 
counterpart in
the electroweak theory--unless one postulates it!. All in all it is quite
natural that the 
non-perturbative QCD effects become relevant at low energy.}.
Thus we see that, 
within the Cosmon context, the correct order of magnitude
for the mass of the necessary scalar can be achieved if one assumes that
dilatation symmetry is spontaneously broken at some intermediate 
GUT scale:
$M_{F}\ll M\ll M_{X}$.  The order of magnitude of $M$ is 
pinned down by the
positivity condition $\beta_{\Lambda}>0$ in eq.(\ref{MasterRGE}), and so
ultimately by the value of the lightest neutrino masses and of the
cosmological constant.

We note that typical intermediate scales $M\sim10^{10}\,GeV$ have been
advocated in the literature \cite{Casas} to generate  a 
suitable RH neutrino
Majorana mass term in the SM that could explain the smallness of the light
neutrino masses using the standard see-saw mechanism in combination with a
Dirac mass term. After all a RH neutrino is a most natural 
object in any gauge
GUT group of rank equal or above that of $SO(10)$. In our framework it is
amusing to imagine that the scale $M$ of SSB of dilatation symmetry could
effectively generate this ``Majorana mass''. To this end we start from the
dilatation symmetric 
formulation of the SM and then extend it by introducing a
gauge and dilatation invariant coupling of the RH neutrinos with that 
scalar
field $\chi$ whose VEV, $M$, is responsible for the SSB of dilatation
symmetry. Then the Yukawa Lagrangian for neutrinos reads
\begin{equation}
\mathcal{L}_{\nu}=\overline{L}
\,\,\,\widetilde{\Phi\,}\mathcal{A\,}\nu_{R}\,\,
+\chi\,\overline{\nu_{R}}^{c}\mathcal{B}\,\nu_{R}+h.c. \label{MMT}
\end{equation}
where $L$ stands for the array of ordinary $SU(2)_{L}$ lepton doublets,
$\widetilde{\Phi\,}$ is the conjugate Higgs doublet, 
$\nu_{R}$ is the array of
RH neutrino fields and 
$\mathcal{A}$, $\mathcal{B}$ are dimensionless mixing
matrices in flavor space. The first operator in (\ref{MMT}) is 
the ordinary
term which furnishes Dirac masses to neutrinos after $\Phi$ spontaneously
breaks the gauge symmetry by acquiring a VEV: 
$\,\left\langle \Phi\right\rangle=v/\sqrt{2}$. 
The second term is the new operator. Obviously, it is invariant
under dilatations. And, as 
it must also respect the gauge symmetry, the scalar
field $\chi$ should be a $SU(2)_{L}\times U(1)_{Y}$ singlet. By writing
$\chi=M\,e^{S/M}$ we recover the dilaton field in the non-linear form
formulated above, and the operator develops a ``Majorana mass 
matrix''
$\mathcal{M}=M\,\mathcal{B}$, 
whose overall scale is determined by $M\gg v$,
basically the mass of the RH neutrinos. 
In this context the scale of SSB of
dilatation symmetry behaves as the RH Majorana mass 
scale, that otherwise must
be introduced by an additional fiat\footnote{A LH Majorana mass term is in
principle another candidate, but then gauge invariance and renormalizability
would require the introduction of a Higgs triplet. However, as already
mentioned, we do not wish 
to alter the Higgs structure of the SM. On the other
hand, it is well-known that one could have a LH Majorana mass term and 
still
preserve 
the SM Higgs structure by introducing a non-renormalizable dimension
$5$ 
operator \cite{Weinberg2} which would induce a mass $\sim v^{2}/M$. This
operator would violate dilatation invariance unless it couples the inverse
power of the field $\chi$. Moreover, for 
$M\sim10^{10}\,GeV$ -- as required in
our framework -- the coefficient of that operator should be 
unnaturally small
to account for the lightest neutrino masses within our double hierarchy
Ansatz.}. Certainly this issue deserves further study. Here we just remark
that, 
for $M$ in the intermediate range $\sim10^{10}\,GeV$ suggested above,
the LH 
neutrinos could acquire the necessary tiny masses by the usual see-saw
mechanism and at the same time the pseudo-dilaton field $S$ receive the
suitable mass to guarantee the right order value, and correct sign, for the
Cosmological Constant.

Interestingly enough, due to the anomaly, the 
linearized effective Lagrangian
contains a residual piece of gravitational strength. After canonical
normalization of the Cosmon kinetic term -- following from (\ref{AWI}) and
(\ref{DilatJ}) --this piece reads
\begin{equation}
\mathcal{L}_{S}=\frac{S}{M\sqrt{1+12h}}\,\Theta_{\mu}^{\mu}\,=f\,\,G^{1/2}
S\,\,\Theta_{\mu}^{\mu}, \label{EL}
\end{equation}
where
\begin{equation}
f=\sqrt{\frac{16\pi h}{1+12h}\,.} \label{fh}
\end{equation}

We emphasize that an important virtue of the Cosmon-dilaton picture is that
this scalar ``completely'' decouples from the SM Lagrangian, as can be 
easily
demonstrated by a conformal transformation of all the matter fields,
$\varphi\rightarrow$ $e^{-D\,S/M}$ $\varphi$ (according to their canonical
dimension $D$) and a Weyl rescaling of the metric tensor: $g_{\mu\nu
}\rightarrow$ $e^{2S/M}g_{\mu\nu}$. In this conformal basis one may check
explicitly 
that the only couplings of $S$ are derivative couplings -- as in
fact 
could be expected from a (pseudo-) Goldstone boson -- with the exception
of the 
anomalous term (\ref{EL}), which in this basis comes out rescaled by a
factor of $e^{-4S/M}$.

Consider 
next the (gravitational-like) macroscopic force mediated by Cosmon
exchange. For a nucleon $N$, the ``Cosmon charge'' 
is given by $f$ times the
nucleonic 
matrix element of the operator (\ref{Thetamumu}). We do not expect
that within a nucleon there is any significant remnant of the 
electroweak and
gravitational anomaly terms. Therefore,
\begin{equation}
Q=f\left\langle N\mid\Theta_{\mu}^{\mu}\mid N\right\rangle =f\left\langle
N\mid\frac{\beta(g_{s})}{2g_{s}}F_{a}^{\mu}F_{\mu}^{a}+m_{q}\gamma
(g_{s})\overline{q}q\mid N\right\rangle , \label{Qcosmon}%
\end{equation}

In the presence of a nucleus we have added in (\ref{Qcosmon}) the anomalous
dimension contribution from the quark mass operator. Thus, formally, we
recover exactly the same expression for the macroscopic force carried 
by the
old Cosmon model \cite{PSW}. 
Except that in the present instance, the Compton
wavelength of this Cosmon is much shorter: $\lambda_{S}=1/m_{S}\gtrsim
(.2-.02)$ mm. Therefore we expect it to mediate a submillimeter range
macroscopic force $\sim\alpha/r^{2}$ supersimposed on the normal 
gravitational
interaction. Furthermore, because it is the anomalous trace of the
energy-momentum tensor -- rather than the trace of the full 
energy-momentum
tensor -- that 
enters eq.(\ref{Qcosmon}), this submillimeter new interaction
is not only of pure gravitational mass-dependent nature, but it 
can also carry
(hierarchically weaker) composition-dependent  (baryon number and isospin
number dependent) components -- similarly as in \cite{PSW}. The relative
strength of the dominant component (the mass-dependent one) of the new
interaction with respect to gravity is given by $\,\alpha= f^{2}/4\pi$. 
Of course 
its value hinges on the value of the parameter $f$ in eq.(\ref{Qcosmon}), 
which in turn depends on the ratio between the scales $M$ and $M_{P}$. 
If the former is, as we already suggested above, an intermediate GUT scale
several orders of magnitude below $M_{X}$, then from eqs.(\ref{vh}) and
(\ref{fh}) we get $\alpha=1/3$ -- to a very good approximation-- and the new
submillimetric interaction is comparable to gravity!.

We remark that, in contradistinction to the approach of \cite{PSW}, 
here we do
not propose any relation between the VEV's of the full 
energy-momentum tensor
and its anomalous part, and so no corresponding dynamical 
adjustment mechanism
has been called upon that could lead to previously described inconsistencies
\cite{Wein1}. The Cosmon field $S$ serves here only to make the case of a
light scalar degree of freedom that could realize the positivity 
condition on
the RHS of eq.(\ref{MasterRGE}) 
while at the same time to evade all known
present bounds on light scalars -- and 
still leave a physical imprint of its
existence!
And all this can in principle be achieved without perturbing the
effective low-energy structure of the SM Lagrangian. Whether the basic
properties of 
the framework that we have described here can be preserved in
minimal extensions of the SM will be addressed elsewhere \cite{SS2}.

\section{Discussion and conclusions}

From the previous considerations we may outline the following heuristic
picture. The reason why there are very light (massive) degrees 
of freedom in
our Universe could be related to the (measured) existence of a very small
residual cosmological constant $\Lambda_{ph}$ in the late epochs of its
history. We 
think of this constant as a running parameter $\Lambda_{ph}(t)$
in a Quantum Field Theory of matter and gravitation. Although 
this theory has
not been clearly uncovered, yet, we have explored the consistency of the
$\Lambda_{ph}=0$ initial condition by applying the RG method to the SM,
coupled to Einstein 
gravity, as the effective low-energy theory. We have shown
that the RG structure of the SM is such that the light degrees 
of freedom do
not lead to a ``runaway value'' of $\Lambda_{ph}$ in neighboring infrared
points. In fact, the presently measured value of $\Lambda_{ph}$ could be
explained in 
terms of these light particles. But of course there are other
difficulties. 
Inherent to this approach is the issue of imposing the original
strong boundary condition 
on the RGE of $\Lambda_{ph}$ at a certain point in
the very far cosmological infrared. We cannot justify 
this matter on first principles, but we take it as a 
natural fact based on the observed smallness
of the measured CC. Notice, as a daring remark, 
that the parameters involved
in the cancellation, being those defined at the very far IR, are in a sense
unobservable. However, what we possibly observe nowadays is the value of
 $\Lambda_{ph}(t)$ -- given  by eq.(\ref{4}) -- at an earlier IR point in
cosmological time, that is, before $\Lambda_{ph}(t)$ ``red-shifts'' 
until its
complete extinction, perhaps attracted by some very far, IR-stable, fixed
point -- corresponding to $\beta_{\Lambda}>0$ in eq.(\ref{MasterRGE}). 
If we
would further entertain the possibility that the strong boundary 
condition at
issue just defines that IR-stable fixed point, then this could explain the
natural flow of $\Lambda_{ph}(t)$ towards $\Lambda_{ph}(0)=0.$

In our approach, far from addressing the details of the fundamental 
underlying
theory, we have studied the phenomenological consequences associated to a
radiative departure from the initial IR condition. To this end, we have
adopted the effective field theoretical assumption that the various
contributions at higher and higher energies depend on the progressive
excitation 
of heavier and heavier degrees of freedom, of masses $M_{\alpha}$,
when $\mu>M_{\alpha}$. The ultimate justification for applying this 
procedure
to $\Lambda_{ph}$ and using only the lightest degrees of freedom is the
reasonable order of the predictions for $\Lambda_{ind}$, and 
finally for the
physical $\Lambda_{ph}$, at the energies far below the electron mass. We
stress that the deviation from this effective method would produce a large
value of the CC incompatible with the Standard Cosmological Model 
(Cf. \cite{SS2}) 
and it could ultimately imply that the traditional definition of
the physical CC given in 
eq.(\ref{4}) is inconsistent. However, at the present
stage of knowledge, 
we prefer to maintain this definition and the effective
theory approach. In view of the nice interconnection that it could lead
between the recent cosmological observations and the last experimental
findings in neutrino physics, we rather think of it as a theoretical 
Ansatz
that could be useful inasmuch as it leads to new, testable, physical
consequences. For example, recently there has been much interest in
short-range macroscopic forces generated by string-like 
models and in general
by models that introduce extra, compact, dimensions at the millimeter scale
\cite{Extradim}. In the last reference it is emphasized that these
 short-range
forces are not in contradiction with existing macroscopic experiments
\cite{gravimeter}. From our line of thought we have found an additional,
though completely different, motivation to look for new gravitational 
forces
in Nature: 
namely, they are associated to the existence of new light degrees
of freedom reflecting a non-zero value of the cosmological constant.

Part of this light spectrum is familiar, like the lightest SM neutrino
(possibly the electron neutrino), and maybe also an accompanying sterile
neutrino of similar mass into which the latter can oscillate. We have
emphasized that this scenario could be consistent with all neutrino
experiments. However, the measured sign $\Lambda_{ph}>0$ could be an
indication that a light scalar $S$ of a mass similar to that of these
neutrinos (in fact a few times larger) should exist in order to tilt the
balance of the predicted cosmological constant into the positive 
range, and at
the same time to secure the sign $\beta_{\Lambda}>0$ that guarantees IR
stability of the RG flow. If the necessary scalar is sterile (as it is the
case for the extra neutrino), then there would be no feasible experimental
method to detect it, and thus no chance to substantiate this approach. 
Indeed,
one can explore various models for the light scalar $S$ with the necessary
magnitude of the mass. For instance, it can be some string-induced dilaton
(see e.g. \cite{DP}) or the anomaly-induced scalar (see e.g. \cite{CS}), 
both
with a mass generated by some dynamical mechanism. In this respect we recall
that all 
versions of string gravity  boil down, in the low-energy limit, to
Einstein gravity plus a new ingredient: the dilaton. This is all the more
significant as long as string theory is the only known framework which 
treats
gravity in a way consistent with quantum mechanics. 
Therefore, on very general
grounds, a dilaton type of scalar $S$ is a quite a generic object in the
resulting Quantum Field Theory below $M_{P}$, and under appropriate
circumstances 
it could effectively behave as a Cosmon of the type described
above. If this would be the case, or if an alternative theoretical 
framework
would lead 
to a non-linear Goldstone-mode realization of dilatation symmetry
below some intermediate unification scale $M$, the low-energy 
structure of the
SM would remain essentially unaltered, but there would appear a new
macroscopic force in Nature, whose ultimate reason could stem from the
observed non-vanishing value of the cosmological constant. Moreover, this
framework could also explain the origin of the tiny neutrino masses by
identifying the intermediate 
scale $M$ of SSB of dilatation symmetry with the
typical scale of the RH Majorana masses. 
The potential detection, in future
experiments \cite{gravimeter}, of 
new gravitational-like forces down
to the submillimeter range -- perhaps being attractive and exhibiting a
composition-dependent nature -- could be the sought-after ``smoking gun''
hinting at the possibility of this whole scenario.

%%%%%%%%%%%%%%%%%%%%%%%%%%%%%%%%%%%%%%%%%%%%%%%%%%%%%%%

\vskip5mm \noindent\textbf{Acknowledgments.} I.L.Sh. is grateful to 
M. Asorey
for useful critical observations and discussions. J.S. would like to 
thank M. Carena, M.J. Herrero and C. Wetterich for helpful conversations; 
and also J. Guasch for pointing out a slip in an 
earlier version.$\,\,$ Authors
acknowledge the warm hospitality 
at the Grup de F\'{\i}sica Te\`{o}rica/IFAE
-- Universitat Aut\`{o}noma de Barcelona (I.L.Sh.) 
and the Departamento de
F\'{\i}sica Te\'{o}rica -- Universidad de Zaragoza (I.L.Sh. and J.S.), 
where
the first discussions on this work started. I.L.Sh. is indebted to the
Departamento de Fisica of the Universidade Federal de Juiz de Fora for
 kind hospitality and to the CNPq (Brazil) for the grant. The work 
of J.S. has been supported by CICYT under project No. AEN98-1093.

%\newpage
%%%%%%%%%%%%%%%%%%%%%%%%%%%%%%%%%%%%%%%%%%%%%%%%%%%%%%%
%
%	

\end{document}